\documentclass{iau}
\usepackage{graphicx}

\title[Homogeneous Studies of Transiting Planets]{HSTEP -- Homogeneous Studies of Transiting Extrasolar Planets}

\author[John Southworth]{John Southworth}

\affiliation{Astrophysics Group, Keele University, Staffordshire, ST5 5BG, UK\\email: {\tt astro.js@keele.ac.uk}}

\pubyear{2013}
\volume{293}  %% insert here IAU Symposium No.
\pagerange{1--6}
\jname{Formation, Detection and Characterization of Extrasolar Habitable Planets}
\editors{N. Haghighipour, ed.}
\begin{document}

\maketitle

\begin{abstract}
I give a summary of the HSTEP project: an effort to calculate the physical properties of the known transiting extrasolar planets using a homogeneous approach. I discuss the motivation for the project, list the 83 planets which have already been studied, run through some important aspects of the methodology, and finish with a synopsis of the results.
\end{abstract}

%%%%%%%%%%%%%%%%%%%%%%%%%%%%%%%%%%%%%%%%%%%%%%%%%%%%%%%%%%%%%%%%%%%%%%%%%%%%%%%%%%%%%%%%%%%%%%%%%%%

\firstsection\section{Introduction}

The HSTEP project has now been running for six years, with the aim of measuring the physical properties of as many transiting extrasolar planets (TEPs) as possible using a single homogeneous methodology. Results have now been published for 83 TEPs (Southworth 2008, 2009, 2010, 2011, 2012; hereafter referred to as Papers I to V) and are compiled into the {\it Transiting Extrasolar Planets Catalogue}\,\footnote{TEPCat (Paper\,IV) is available at: {\tt http://www.astro.keele.ac.uk/jkt/tepcat/}}) TEPCat.

The primary motivation for this project was to enable detailed statistical studies of the known TEPs, by removing differences in analysis which made published results difficult to compare directly. A secondary stimulus was the realisation that many results of individual TEPs could be improved with a more extensive or detailed analysis. My inspiration for much of the HSTEP methodology stems from the large body of work on eclipsing binary stars -- of which a TEP system is a special case -- for which mathematical models and results have been available for a century or more (e.g.\ Stebbins 1911; Russell 1912).

Other people have also had this idea. The HAT (Bakos et al., 2004) and SuperWASP (Pollacco et al., 2006) surveys both analyse their newly discovered planets using standard computer codes, although these have gradually evolved over the lifetimes of the projects, whilst Torres, Winn \& Holman (2008) produced an excellent homogeneous analysis of 23 TEPs. An important area where standardisation is needed is the recovery of the host stars' atmospheric parameters ($T_{\rm eff}$, $\log g$ and $[{\rm Fe/H}]$) from high-dispersion spectra; existing work in this area (Ammler-von Eiff et al.\ 2009; Torres et al.\ 2012; Doyle et al.\ 2012) either suffers from systematic errors or considers only a small number of stars. Their results are nevertheless incorporated into HSTEP when appropriate.

The full list of objects studied as part of the HSTEP project is given in Table\,1. The gory details of their analysis can be found in HSTEP papers so will not be repeated here. Instead, I use this conference proceedings to provide an index to the HSTEP approach of deriving the physical properties of a TEP and its host star. The index is arranged not alphabetically, but in sequential order for undertaking such an analysis. It is not complete, but hopefully is also not completely useless.

\begin{table} \begin{center}
\caption{TEP systems included in HSTEP. The lower part of the table gives the TEPs
which are the subject of dedicated papers presenting new transit light curves.}
\begin{tabular}{p{1.9cm}p{1.9cm}p{1.9cm}p{2.2cm}p{2.4cm}p{1.8cm}} \hline
CoRoT-1  & CoRoT-14 & HAT-P-7    & Kepler-8   & OGLE-TR-56  & WASP-10 \\
CoRoT-2  & CoRoT-15 & HAT-P-9    & Kepler-12  & OGLE-TR-111 & WASP-12 \\
CoRoT-3  & CoRoT-17 & HAT-P-11   & Kepler-14  & OGLE-TR-113 & WASP-13 \\
CoRoT-4  & CoRoT-18 & HAT-P-14   & Kepler-15  & OGLE-TR-132 & WASP-14 \\
CoRoT-5  & CoRoT-19 & HD\,17156  & Kepler-17  & OGLE-TR-182 & WASP-21 \\
CoRoT-6  & CoRoT-20 & HD\,80606  & KOI-135    & OGLE-TR-211 & XO-1    \\
CoRoT-7  & CoRoT-23 & HD\,149026 & KOI-196    & OGLE-TR-L9  & XO-2    \\
CoRoT-8  & GJ\,436  & HD\,189733 & KOI-204    & TrES-1      & XO-3    \\
CoRoT-9  & HAT-P-1  & HD\,209458 & KOI-254    & TrES-2      & XO-4    \\
CoRoT-10 & HAT-P-2  & Kepler-4   & KOI-423    & TrES-3      & XO-5    \\
CoRoT-11 & HAT-P-3  & Kepler-5   & KOI-428    & TrES-4      &         \\
CoRoT-12 & HAT-P-4  & Kepler-6   & LHS\,6343  & WASP-1      &         \\
CoRoT-13 & HAT-P-6  & Kepler-7   & OGLE-TR-10 & WASP-3      &         \\
\hline
\ \ HAT-P-5  & \multicolumn{2}{l}{Southworth et al.\ (2012b)} & \ \ WASP-5   & \multicolumn{2}{l}{Southworth et al.\ (2009a)} \\
\ \ HAT-P-13 & \multicolumn{2}{l}{Southworth et al.\ (2012a)} & \ \ WASP-7   & \multicolumn{2}{l}{Southworth et al.\ (2011) } \\
\ \ WASP-2   & \multicolumn{2}{l}{Southworth et al.\ (2010) } & \ \ WASP-17  & \multicolumn{2}{l}{Southworth et al.\ (2012c)} \\
\ \ WASP-4   & \multicolumn{2}{l}{Southworth et al.\ (2009b)} & \ \ WASP-18  & \multicolumn{2}{l}{Southworth et al.\ (2009c)} \\
\hline \end{tabular} \end{center} \end{table}

%%%%%%%%%%%%%%%%%%%%%%%%%%%%%%%%%%%%%%%%%%%%%%%%%%%%%%%%%%%%%%%%%%%%%%%%%%%%%%%%%%%%%%%%%%%%%%%%%%%

\section{Step 1: light curve analysis}

{\bf Data} \ \ The observational data used in the HSTEP project consists purely of light curves of transit events. In most cases I use published data, but for some TEPs I obtain new photometry and present it in dedicated papers (see Table\,1). In many cases, particularly for {\it Kepler} and CoRoT data, I fit a polynomial to rectify each transit to the continuum, before analysing the data from multiple transits simultaneously.

{\bf Geometrical model} \ \ The transit light curves are fitted using the {\sc jktebop} code\,\footnote{{\sc jktebop} is available at: {\tt http://www.astro.keele.ac.uk/jkt/codes/jktebop.html}} (Southworth, Maxted \& Smalley 2004a; Southworth, Bruntt \& Buzasi 2007), which represents the star and planet as spheres for the calculation of eclipses and as biaxial spheroids for calculating proximity effects.

{\bf Limb darkening} \ \ LD is a tricky phenomenon but can be easily incorporated using parametric functions (``laws''). The linear LD law has one coefficient and represents the LD of the Sun very poorly; it is inadequate for satellite-quality data but usually good enough for ground-based photometry (Paper\,I). Several two-coefficient laws exist, and these are sufficiently precise for all existing transit data. In HSTEP I use the linear law to establish context, but base all results on solutions using four two-coefficient laws (quadratic, logarithmic, square-root and cubic).

{\bf Limb darkening coefficients} \ \ Extremely good data are required to fit for both coefficients of an LD law, because the two coefficients are strongly correlated (e.g.\ Southworth, Bruntt \& Buzasi 2007). In most cases I therefore fix one or both of the coefficients to theoretical values obtained via stellar model atmospheres. I use the {\sc jktld} code\,\footnote{{\sc jktld} is available at: {\tt http://www.astro.keele.ac.uk/jkt/codes/jktld.html}} to interpolate tabulated predictions to the required $T_{\rm eff}$ and $\log g$ of the star. Theoretical LDCs are imperfect, as can be seen by comparing different studies, so their uncertainty is taken into account by perturbing fixed LDCs during the error analysis.

{\bf Contaminating light} \ \ Third light from a faint nearby object dilutes the variations in brightness of a TEP system. This reduces the observed transit depth, causing an underestimate of the radius of the planet. A detailed investigation of this (Paper\,III) showed that a third light of 5\% decreases the measured radius of the planet by 2\%. Third light cannot just be included as a fitted parameter as it is almost perfectly correlated with other parameters (Paper\,III). It can however be accounted for, if its value is known. This is done by making third light a free parameter in the {\sc jktebop} analysis, constrained by the observed value, which also allows incorporation of the uncertainty.

{\bf Orbital eccentricity} \ \ This affects the orbital speed of the planet and therefore modifies the duration of the transit. Kipping (2008) showed that eccentricity cannot be derived from transits themselves because its effect on their {\it shape} is extremely subtle. Eccentricity, $e$, and longitude of periastron, $\omega$, must therefore be incorporated as constraints in the same way as for third light (see Paper\,III). This option is allowed for in {\sc jktebop}. When possible I use the combination terms $e\cos\omega$ and $e\sin\omega$, as these are much less strongly correlated than $e$ and $\omega$ themselves.

{\bf Numerical integration} \ \ The long integration times of the {\it Kepler} and CoRoT satellites (1765\,s and 512\,s, respectively, for their long-cadence targets) smears out the shape of transits. This can be accounted for by integrating the model by the correct amount prior to comparing it to the observational data. In Paper\,IV I showed that the effect of ignoring numerical integration can be huge for {\it Kepler} long-cadence data. Kipping (2010) has studied this effect in more detail.

{\bf Statistical errorbars} \ \ Monte Carlo simulations (Southworth et al., 2004b) are used to assign statistical errorbars to the quantities derived from the transit light curves. I do not incorporate a Markov chain -- exploration of the parameter space is instead initiated by perturbing the initial parameters used when fitting each synthetic light curve generated as part of the Monte Carlo algorithm.

{\bf Systematic errorbars} \ \ Correlated noise afflicts all ground-based light curves, in effect lowering the information content of the data by violating the {\it `iid'} (independent and identically distributed) ideal. It can dominate the statistical noise in the data, and has a more deleterious effect on the results (see e.g.\ Pont, Zucker \& Queloz 2006; Carter \& Winn 2009). I deal with this using a residual-permutation algorithm (Paper\,I) and adopt the larger of the statistical or systematic errorbar for each output parameter.

{\bf Multiple light curves} \ \ If several transit light curves exist for the same TEP, I model them independently and combine the results into weighted means. This make any disagreements between different datasets obvious and quantifiable, allowing the errorbars to be increased to reflect the situation. This is a major advantage over the usual simultaneous solutions, which require great care to tease out such discrepancies.

%%%%%%%%%%%%%%%%%%%%%%%%%%%%%%%%%%%%%%%%%%%%%%%%%%%%%%%%%%%%%%%%%%%%%%%%%%%%%%%%%%%%%%%%%%%%%%%%%%%

\section{Step 2: physical properties}

{\bf Input quantities} \ \ The results from analysing the light curves are now augmented with published values for $T_{\rm eff}$, $\log g$, $[{\rm Fe/H}]$ and the velocity amplitude of the star induced by the orbiting planet. If this were an eclipsing binary with both stars visible in the spectrum, it would now be possible to calculate their masses and radii from just the light curve results and the two velocity amplitudes. Because the velocity amplitude of the planet is inaccessible, we have insufficient information to do this for TEP systems. We require an {\it additional constraint} to plug the gap.

{\bf Additional constraint from stellar models} \ \ Option 1 for this {\it additional constraint} is to use the predictions of stellar models to provide essentially a mass--radius relation for the host star. The star's density turns out to be accurately determined from the transit light curves alone (Seager \& Mall{\'e}n-Ornelas 2003); providing an alternate relation from stellar theory allows density ($M$/$R^3$) to be transformed into $M$ and $R$ separately.

{\bf Additional constraint from eclipsing binary stars} \ \ Option 2 for the {\it additional constraint} is a calibration of stellar properties based on stars with known mass and radius, from their presence in an eclipsing binary system. I introduced this concept in Paper\,II, adopting a simple mass--radius relation which has since turned out to be too crude due to its neglect of stellar evolution. Torres, Andersen \& Gim{\'e}nez (2010) proposed multi-parametric relations for $M$ and $R$ from $T_{\rm eff}$, $\log g$ and $[{\rm Fe/H}]$. Enoch et al.\ (2010) modified these to work with density instead of $\log g$, as density is quantifiable directly from transit light curves. In Paper\,IV, I derived an improved calibration using more stars (180 versus Enoch's 38), and ignoring masses above 3\,M$_\odot$ (Enoch's calibration extended to 15\,M$_\odot$) as these are irrelevant for typical TEP host stars. I use this calibration to provide an almost-empirical {\it additional constraint}.

{\bf Calculating physical properties} \ \ I use a code originally written for eclipsing binary stars, {\sc jktabsdim}, to calculate the physical properties of the star and planet from the already-measured photometric and spectroscopic parameters. The genealogy of the code allowed me to adopt the approach of using the (unknown) velocity amplitude of the secondary component (i.e.\ the planet) as a control parameter, which is iteratively refined to match the {\it additional constraint}. The figure of merit for this is basically the agreement between the observed and predicted $T_{\rm eff}$, and the calculated and predicted stellar radius, for the known $T_{\rm eff}$ and $[{\rm Fe/H}]$ and the calculated mass of the star. The {\it additional constraint} is therefore encoded completely and exclusively in the velocity amplitude of the planet. In Paper\,II, I used three different sets of stellar models for the  {\it additional constraint}, and this was revised to five sets of models beginning in Paper\,III.

{\bf Statistical uncertainties} \ \ {\sc jktabsdim} adopts a perturbation approach (Southworth, Maxted \& Smalley 2005) to propagate uncertainties through the solution process described above. This is done by shifting every input parameter by its errorbar and calculating a new solution, from which I measure the error in each output parameter due to the error in this input parameter. The overall errorbar for each output parameter is simply the quadrature addition of the ones from all input parameters, leading not just to robust errorbars but also a complete error budget for the whole process.

{\bf Systematic errors} \ \ The use of theoretical stellar models to provide the {\it additional constraint} induces systematic errors in the final results: the models are not perfect. Crucially, my use of multiple sets of model predictions means that the strength of the agreement between models can be quantified. I am therefore able to provide separate statistical and systematic errorbars for every output parameter calculated using {\sc jktabsdim}. The systematic errors are generally fairly mild: 1\% for stellar mass, 0.6\% for planetary mass, and 0.3\% or less for other properties (Paper\,III). Some quantities are free of systematics: the planetary surface gravity can be obtained from only the light curve and radial velocities (Southworth, Wheatley \& Sams 2007); the planetary equilibrium temperature is not affected by the theoretical models (Paper\,III); the stellar density is only negligibly impacted by the models (Seager \& Mall{\'e}n-Ornelas 2003).

{\bf Physical constants} \ \ {\sc jktabsdim} requires a set of physical constants, specifically $\pi$, $\sigma_{\rm SB}$, the AU, $GM_\odot$, $G$, $R_\odot$, the ratio $M_\odot/M_{\rm Jup}$ and $R_{\rm Jup}$. Revisions to these constants are expected in the near future to account for the discrepancy between the equatorial radius of Jupiter (widely adopted to represent $R_{\rm Jup}$) and the volume-equivalent radius (see Paper\,IV). The physical constants used in HSTEP are given in Paper\,IV and TEPCat.

{\bf Previous work} \ \ I make extensive comparisons to published physical properties, in order to confirm the reliability of my results, to track how the publication of new data improves the measurements, and to find mistakes in published papers (in several cases I have been able to repair numbers which have previously been calculated incorrectly). In Paper\,III, I also compared the results from the discovery papers of WASP-11 / HAT-P-10, a planet which was independently identified by SuperWASP and HATNet, providing a fortuitous check on the interagreement of analyses by different research groups. I found that the agreement was generally good except for two parameters: the ratio of the radii of the components and the stellar $T_{\rm eff}$. This confirmed a gut feeling that these two parameters are more affected by systematics than is generally appreciated.

%%%%%%%%%%%%%%%%%%%%%%%%%%%%%%%%%%%%%%%%%%%%%%%%%%%%%%%%%%%%%%%%%%%%%%%%%%%%%%%%%%%%%%%%%%%%%%%%%%%

\section{Results}

\begin{figure}[t]
\includegraphics[width=\textwidth,angle=0]{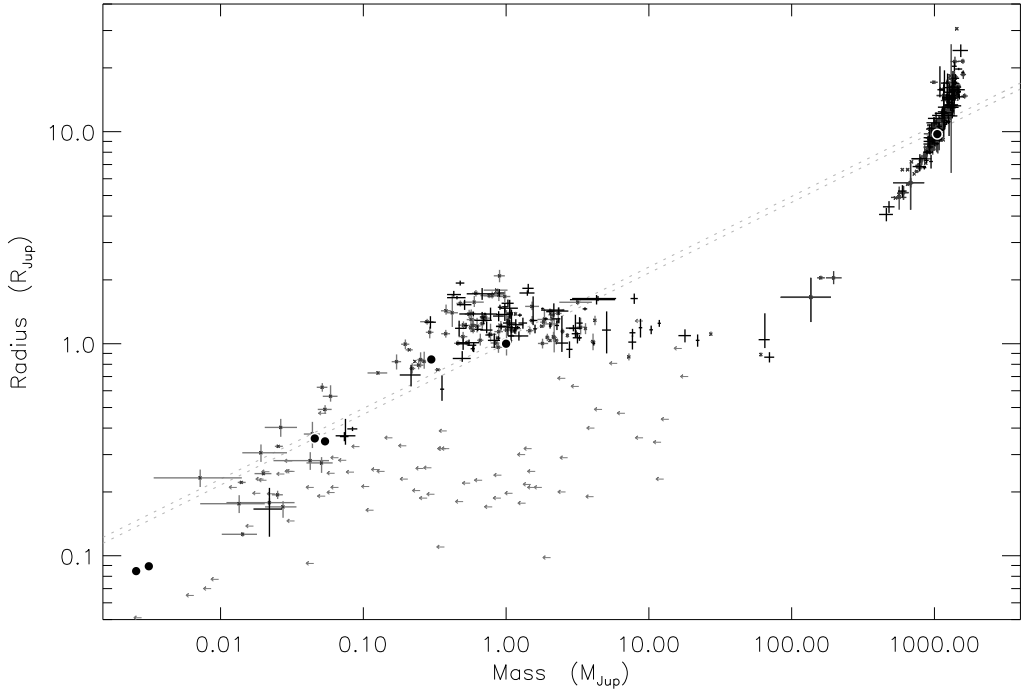}
\caption{Plot of the masses versus the radii of TEPs and their host stars, as of November 2012.
Black crosses show the objects studied in the HSTEP papers and grey crosses show results
taken from the literature. The Solar system bodies are shown by black filled circles.
The grey dotted lines show the loci of densities equal to those of Jupiter and the Sun.
TEPs with only an upper limit on their mass are shown with left-pointing arrows.}
\end{figure}

The total number of TEPs studied (including the papers in Table\,1): 14 in Paper\,I (light curves only), 14 in Paper\,II (full analysis), 30 in Paper\,III, 58 in Paper\,IV and now 83 in Paper\,V. The results are compiled in TEPCat. Future work is planned to push the number well into triple figures. This compares unfavourably with the 270 systems currently known (see TEPCat); the bottleneck for the HSTEP project is the available manpower. Fig.\,1 summarises the masses and radii of the host stars and their planets.

My results generally confirm existing values, but there are many exceptions (e.g.\ CoRoT-8 in Paper\,IV, where I found that even the orbital ephemeris was wrong, and Kepler-15 and OGLE-TR-56 in Paper\,V). I also usually arrive at significantly larger errorbars than found in previous works. I often agree with these works to within my errorbars but not their errorbars, suggesting that my errorbars are more reliable.

Correlations exist between orbital period, planet mass and surface gravity (Mazeh, Zucker \& Pont 2005; Southworth et al.\ 2007b). These were found to be of marginal significance in Papers II and III, but by Paper\,V both correlations had become significant at the $4\sigma$ level. A posited connection between planetary equilibrium temperature and Safronov number largely vanished as more TEPs were found (see Paper\,V).

The error budgets I derive are useful for deciding which TEPs would benefit from what kind of further observations. As a general rule the quality of the light curve dominates the precision of the final results, but for some planets (particularly those observed by HST and {\it Kepler}) the systematic errors are becoming critical. One under-appreciated point is that the ephemerides of some TEPs are imprecise, and in some cases are already of limited use for predicting when transits will occur. The main offenders are the newer CoRoT planets, as in many cases the CoRoT light curve covers only a short time interval (e.g.\ CoRoT-14, CoRoT-17 and CoRoT-20, but also CoRoT-4). Further observations of these objects are needed very soon, or we risk losing the transits within a few years.

Finally, extensive plots of results are available in the HSTEP papers, concentrating primarily on the physical properties of the systems but also concerning their sky positions and the discovery rate of these fascinating objects (see Paper\,V).

%%%%%%%%%%%%%%%%%%%%%%%%%%%%%%%%%%%%%%%%%%%%%%%%%%%%%%%%%%%%%%%%%%%%%%%%%%%%%%%%%%%%%%%%%%%%%%%%%%%

%%%%%%%%%%%%%%%%%%%%%%%%%%%%%%%%%%%%%%%%%%%%%%%%%%%%%%%%%%%%%%%%%%%%%%%%%%%%%%%%%%%%%%%%%%%%%%%%%%%

\end{document}